\documentclass[12pt,preprint]{emulateapj}

\usepackage{amssymb}
\usepackage{times}
\usepackage{amsmath}
\usepackage{multirow}
\usepackage[colorlinks,backref,dvipdfm,linkcolor=blue,anchorcolor=,citecolor=blue]{hyperref}
\makeatletter

\newcommand{\teff}{$T_{\mathrm{eff}}$}

\newcommand{\numax}{$\nu_{\mathrm{max}}$}
\newcommand{\dnu}{$\Delta\nu$}
\newcommand{\kepler}{\textit{Kepler}}

\newcommand{\keplermission}{\textit{Kepler Mission}}

\newcommand{\dm}{${\rm(m-M)_{0}}$}

\newcommand{\rmnum}[1]{\romannumeral #1}
\newcommand{\Rmnum}[1]{\expandafter\@slowromancap\romannumeral #1@}
\makeatother

\slugcomment{}

\shorttitle{Asteroseismic Study on Distance Moduli with a New Relationship}
\shortauthors{Wu, Li, \& Hekker}

\begin{document}


\title{Asteroseismic Study on Cluster Distance Moduli for RGB Stars in NGC~6791 and NGC~6819}


\author{
T. Wu\altaffilmark{1,2,3}, Y. Li \altaffilmark{1,2}, and S. Hekker\altaffilmark{4}
}


\altaffiltext{1}{Yunnan Observatories, Chinese Academy of Sciences, P.O. Box 110, Kunming 650011, China; wutao@ynao.ac.cn, ly@ynao.ac.cn, hekker@mps.mpg.de}
\altaffiltext{2}{Key Laboratory for Structure and Evolution of Celestial Objects, Chinese Academy of Sciences, P.O. Box 110, Kunming 650011, China}
\altaffiltext{3}{University of Chinese Academy of Sciences, Beijing 100039, China}
\altaffiltext{4}{Max Planck Institute for Solar System Research, Justus von Liebig Weg 3, 37077 G\"ottingen, Germany}


\begin{abstract}

  Stellar distance is an important basic parameter in stellar astrophysics. Stars in a cluster are thought to be formed coevally from the same interstellar cloud of gas and dust. They are therefore expected to have common properties. These common properties strengthen our ability to constrain theoretical models and/or to determine fundamental parameters, such as stellar mass, metal fraction, and distance when tested against an ensemble of cluster stars.
  Here we derive a new relation based on solar-like oscillations, photometric observations, and the theory of stellar structure and evolution of red giant branch stars to determine cluster distance moduli through the global oscillation parameters \dnu\ and \numax\, and photometric data \textit{V}. The values of \dnu\ and \numax\ are derived from \kepler\ observations. At the same time, it is used to interpret the trends between \textit{V} and \dnu.
  From the analyses of this newly derived relation and observational data of NGC~6791 and NGC~6819 we devise a method in which all stars in a cluster are regarded as one entity to determine the cluster distance modulus. This approach fully reflects the characteristic of member stars in a cluster as a natural sample. From this method we derive true distance moduli of $13.09\pm0.10$ mag for NGC~6791 and $11.88\pm0.14$ mag for NGC~6819. Additionally, we find that the distance modulus only slightly depends on the metallicity [Fe/H] in the new relation. A change of 0.1 dex in [Fe/H] will lead to a change of 0.06 mag in the distance modulus. 

\end{abstract}



\section{Introduction}\label{sec1}
Asteroseismology provides a powerful tool to probe detailed information regarding the internal structure and evolutionary state of stars. Many stars with solar-like oscillation have been observed with space-based instruments, such as \textit{WIRE} \citep[e.g.][]{Hacking99,Buzasi00}, \textit{MOST} \citep[e.g.][]{Walker03,Matthews04}, \textit{CoRoT} \citep[e.g.][]{Baglin06}, and \kepler\ \citep[e.g.][]{Koch10,Gilliland10}. These missions have provided precise near-uninterrupted photometric timeseries data which allows for asteroseismic analyses of many stars. This opens the possibility to study large samples of stars, i.e., to perform so-called ``ensemble asteroseismology'' \citep{Chaplin11}. The observed oscillation parameters can be used to determine the stellar fundamental parameters (mass $M$, radius $R$, surface gravity $g$, mean density $\rho$, etc.).

The members of a cluster constitute a natural sample, as stars in a cluster are assumed to be formed coevally from the same interstellar cloud of gas and dust. Therefore, they are expected to have common properties, such as element composition, distance, age, etc. For this reason, ensemble asteroseismology is very suitable for cluster stars, for examples, see \citet{Stello10,Stello11a,Stello11b}, \citet{Hekker11b}, \citet{basu11}, \citet{Miglio12}, \citet{Corsaro12}, and \citet[][]{Wu13}.

Distance is a fundamental parameter in astrophysics. The \textit{Hipparcos} satellite \citep[e.g.][]{Perryman_ESA97} provided parallax measurements of a large number of stars to obtain their distances. For clusters, there are many methods to obtain the cluster distance modulus or distance. For example,  isochrone fitting \citep[e.g.][]{Chaboyer99,Stetson03,Bedin05,Bedin08,Hole09,Wu13}, or using red-clump stars as ``standard candles'' \citep[e.g.][]{Garnavich94,Gao12}. Additionally, the cluster distance can be derived from a detailed analysis of binary systems \citep[e.g.][]{Brogaard11,Jeffries13,Sandquist13}, from the period-luminosity relation of pulsating stars \citep[e.g.][]{Soszynski08,Soszynski10}, or from direct estimates \citep[e.g.][]{basu11,Miglio12,Balona13}, and so on.


\renewcommand{\arraystretch}{1.}
\begin{deluxetable*}{lccccccr}
\tablewidth{\textwidth}
\tablecaption{Literature overview of cluster distance moduli of NGC~6791 and NGC~6819.\label{table_liter}}
\tablehead{
\colhead{${\rm (m-M)_{0}}$} & \colhead{${\rm (m-M)_{V}}$} & \colhead{$E(B-V)$}  & \colhead{$A_{V}$}  & \colhead{Metallicity\tablenotemark{a,b}} & \colhead{Age} & \colhead{Methods} &  \colhead{Ref.} \\
\colhead{[mag]} & \colhead{[mag]} & \colhead{[mag]} & \colhead{[mag]} & \colhead{$Z$\tablenotemark{a} or [Fe/H]\tablenotemark{b}}& \colhead{[Gyr]}& &
}
\startdata
   \multicolumn{8}{c}{NGC 6791} \\
\hline
   13.55  & 14.21\tablenotemark{aa} &  0.22$\pm$0.02 & 0.66 & 0.01\tablenotemark{a} & \nodata & main-sequence stars & \citet[][]{Kinman65} \\ 
   12.88$\pm$0.6\tablenotemark{aa}   & 13.3$\pm$0.6 &  0.13 & 0.42 & \nodata & \nodata & spectroscopic parallaxes & \citet[][]{Harris_Canterna81} \\ 
   13.48$\pm$0.35\tablenotemark{aa}   & 13.9$\pm$0.35 & 0.13 & 0.42 & \nodata & \nodata & sed-clump stars & \citet[][]{Harris_Canterna81} \\ 
   13.58$\pm$0.2\tablenotemark{aa}   & 14.0$\pm$0.2 &  0.13 & 0.42 & 0.02\tablenotemark{a,c} & $\sim$7 & isochrone & \citet[][]{Harris_Canterna81} \\ 
   13.25 & \nodata & \nodata & \nodata & \nodata & \nodata & red-clump stars & \citet[][]{Anthony-Twarog84} \\
   12.8\tablenotemark{d} & 13.5 & 0.20 & 0.70 & 0.019\tablenotemark{a,c} & 6.0$\pm$0.7& isochrone & \citet[][]{Anthony-Twarog_Twarog85} \\ 
   12.5\tablenotemark{e} & 13.2 & 0.20 & 0.70 & 0.0169\tablenotemark{a,c}& 12.0 & isochrone & \citet[][]{Anthony-Twarog_Twarog85} \\ 
   12.75\tablenotemark{aa,e} & 13.45 & 0.225 & 0.70\tablenotemark{aa} & 0.0169\tablenotemark{a,c}& 10$\sim$12.5 & isochrone & \citet[][]{Kaluzny90} \\ 
   \nodata &   13.65  &   \nodata & \nodata & 0.0\tablenotemark{b,c}& $\sim$9 & red-clump stars & \citet[][]{Zurek93} \\ 
   \nodata  & 13.6& \nodata & \nodata & $-$0.04$\pm$0.12\tablenotemark{b} & $\sim$9 & red-clump stars & \citet[][]{Garnavich94} \\ 
   \nodata  & 13.55& 0.19$\pm$0.03 & \nodata & 0.03\tablenotemark{a} & $\sim$9 & isochrone & \citet[][]{Garnavich94} \\ 
   12.66  &   12.96  &   0.10$\pm$0.02 & \nodata & $+$0.19\tablenotemark{b}& 10 & isochrone & \citet[][]{Montgomery94} \\ 
   12.97 & 13.52& 0.17 & \nodata & $+$0.3\tablenotemark{b} & 7.2 & red-clump stars & \citet[][]{Kaluzny95} \\ 
   12.75$\sim$12.82 & 13.30$\sim$13.37& 0.17 &\nodata & $+$0.2\tablenotemark{b} & 7.2 & main-sequence stars & \citet[][]{Kaluzny95} \\ 
   12.86$\sim$12.93 & 13.41$\sim$13.48& 0.17 &\nodata & $+$0.3\tablenotemark{b} & 7.2 & main-sequence stars & \citet[][]{Kaluzny95} \\ 
   \nodata & 13.49$\sim$13.70 & 0.19$\sim$0.24 & \nodata &$+$0.35\tablenotemark{b} &10$\pm$0.5 & red-clump stars & \citet[][]{Tripicco95} \\ 
   \nodata & 13.49$\sim$13.52 & 0.20$\sim$0.23 & \nodata &$+$0.15\tablenotemark{b} &10 & isochrone & \citet[][]{Tripicco95} \\ 
   \nodata & 13.30$\sim$13.45 & 0.08$\sim$0.13 & \nodata & $+$0.4\tablenotemark{b}&8$\pm$0.5 & isochrone & \citet[][]{Chaboyer99} \\ 
   \nodata & 13.42 & 0.10$\sim$0.11 & \nodata & $+$0.4\tablenotemark{b}&8 & isochrone & \citet[][]{Liebert99} \\ 
  \nodata & $\sim$13.0 & 0.1 & \nodata & \nodata & \nodata & binaries & \citet[][]{Mochejska03} \\ 
  12.79 & \nodata & 0.09 & \nodata & $+$0.3\tablenotemark{b} & 12 & isochrone & \citet[][]{Stetson03} \\ 
  13.0 & 13.5 & 0.15 & \nodata & 0.03\tablenotemark{a} & 9 & isochrone & \citet[][]{King05} \\ 
  13.07$\pm$0.04 & \nodata & 0.14$\pm$0.04 & \nodata & $+$0.4$\pm$0.01\tablenotemark{b} & 8 & red-clump stars & \citet[][]{Carney05} \\ 
  12.93 & \nodata & 0.17 & \nodata & $+$0.3\tablenotemark{b} & 8 & isochrone & \citet[][]{Carney05} \\ 
  12.96 & \nodata & 0.13 & \nodata & $+$0.4\tablenotemark{b} & 8 & isochrone & \citet[][]{Carney05} \\ 
  13.11 & \nodata & 0.11 & \nodata & $+$0.5\tablenotemark{b} & 7.5 & isochrone & \citet[][]{Carney05} \\ 
  13.07$\pm$0.05 & 13.45 & 0.09$\pm$0.01 & \nodata & 0.046\tablenotemark{a} & 8.0$\pm$1.0 & isochrone & \citet[][]{Carraro06} \\ 
  \nodata & 13.35 & 0.13 & \nodata & 0.04\tablenotemark{a} & 8$\sim$9 & isochrone & \citet[][]{Carraro06} \\ 
  13.14$\pm$0.15\tablenotemark{aa}& 13.60$\pm$0.15& 0.15 & 0.46\tablenotemark{aa} & $+$0.45\tablenotemark{b} & 7.0$\pm$1.0 & isochrone & \citet[][]{Anthony-Twarog07} \\ 
  13.0 & \nodata& 0.14 & \nodata & $+$0.37\tablenotemark{b} & 8.5 & isochrone & \citet[][]{Kalirai07} \\ 
  \nodata & 13.30$\pm$0.2 & 0.09 & \nodata & \nodata & \nodata & binaries & \citet[][]{de-Marchi07} \\ 
  13.0 & \nodata & 0.15$\pm$0.02 & \nodata & $+$0.40$\pm$0.10\tablenotemark{b} & 6.2$\sim$9.0 & binary & \citet[][]{Grundahl08} \\ 
  \nodata & 13.46 & 0.15 & \nodata & $+$0.40\tablenotemark{b} & 7.7$\sim$9.0 & isochrone & \citet[][]{Grundahl08} \\ 
  \nodata & 13.51$\pm$0.06 & 0.160$\pm$0.025 & \nodata & $+$0.29$\pm$0.10\tablenotemark{b} & \nodata & binaries & \citet[][]{Brogaard11} \\ 
  13.11$\pm$0.06 &13.61$\pm$0.06\tablenotemark{aa}  & 0.16 & 0.50\tablenotemark{aa} & $+$0.29\tablenotemark{b} & 6.8$\sim$8.6 & asteroseismology & \citet[][]{basu11} \\ 
  13.01$\pm$0.07\tablenotemark{aa} & 13.51$\pm$0.02 & 0.16$\pm$0.02 & 0.50$\pm$0.06\tablenotemark{aa} & $+$0.3\tablenotemark{b} & \nodata & asteroseismology & \citet[][]{Miglio12} \\ 
  12.97$\pm$0.05\tablenotemark{aa} & 13.36$\pm$0.04 & 0.14$\pm$0.01 & 0.43$\pm$0.03\tablenotemark{aa} & 0.04$\pm$0.005\tablenotemark{a} & 8.0$\pm$0.4 & isochrone & \citet[][]{Wu13} \\ 
  \textbf{13.08$\pm$0.08} & \textbf{13.58$\pm$0.03} & \textbf{0.16$\pm$0.025} & \textbf{\nodata} & \textbf{$+$0.29$\pm$0.10\tablenotemark{b}} & \textbf{\nodata} & \textbf{asteroseismology} & \textbf{The present work\tablenotemark{bb}} \\
  \textbf{13.09$\pm$0.10} & \textbf{13.59$\pm$0.06} & \textbf{0.16$\pm$0.025} & \textbf{\nodata} & \textbf{$+$0.29$\pm$0.10\tablenotemark{b}} & \textbf{\nodata} & \textbf{asteroseismology} & \textbf{The present work\tablenotemark{cc}} \\

\hline 
   \multicolumn{8}{c}{NGC 6819} \\
\hline
  11.54  & 11.9 & 0.12 & 0.36 & \nodata & \nodata & main-sequence turnoff & \citet[][]{Burkhead71} \\ 
  11.5   & 12.6 &  0.3 & 0.9  & \nodata & 2 & main-sequence stars & \citet[][]{Lindoff72} \\ 
  11.76  & 12.50 &  0.28 & \nodata & \nodata & \nodata & main-sequence stars & \citet[][]{Auner74} \\ 
  \nodata  & 12.35& 0.16 & \nodata & -0.10$\sim$0.0\tablenotemark{b} & 2.4 & isochrone/ZAHB & \citet[][]{rv98} \\ 
  \nodata  & 12.30$\pm$0.12& 0.10 & \nodata & 0.02\tablenotemark{a} & 2.5 & isochrone  & \citet[][]{Kalirai01} \\ 
  \nodata  & 12.30& 0.10 & \nodata & 0.019\tablenotemark{a,c} & 2.4 & isochrone & \citet[][]{Hole09} \\ 
  \nodata  & 12.38& \nodata & \nodata & \nodata & \nodata & binary & \citet[][]{Talamantes10} \\ 
  11.85$\pm$0.05& 12.31$\pm$0.05\tablenotemark{aa} & 0.15 & 0.46\tablenotemark{aa} & $+$0.09\tablenotemark{b} & 2$\sim$2.4 & asteroseismology & \citet[][]{basu11} \\ 
  11.34$\pm$0.02\tablenotemark{aa}& 11.80$\pm$0.02 & 0.15 & 0.46\tablenotemark{aa} & 0.0\tablenotemark{b} & \nodata & asteroseismology & \citet[][]{Miglio12} \\ 
  \nodata  & 12.50& 0.14 & \nodata & $+$0.09\tablenotemark{b} & 2.25 & isochrone & \citet[][]{Anthony-Twarog13} \\ 
   \nodata & 12.39$\pm$0.08 & \nodata & \nodata & $+$0.09\tablenotemark{b} & 2.65$\pm$0.25 & binaries & \citet[][]{Sandquist13} \\ 
  12.00$\pm$0.05  & 12.37$\pm$0.10 & 0.12$\pm$0.03 & \nodata & \nodata & \nodata & dwarf stars near the turnoff & \citet[][]{Jeffries13} \\ 
   \nodata & 12.28$\sim$12.40 & 0.12$\pm$0.03 & \nodata & $+$0.06$\sim$$+$0.13\tablenotemark{b} & 2.1$\sim$2.5 & isochrone & \citet[][]{Jeffries13} \\ 
   \nodata& 12.44$\pm$0.07 & \nodata & \nodata & $+$0.09$\pm$0.03\tablenotemark{b} & 2.2$\sim$3.7 & binaries & \citet[][]{Jeffries13} \\ 
   11.88$\pm$0.08 & 12.34$\pm$0.08\tablenotemark{aa} & 0.15 & 0.46\tablenotemark{aa} & & \nodata & asteroseismology & \citet[][]{Balona13} \\ 
   11.94$\pm$0.04 & 12.40$\pm$0.04\tablenotemark{aa} & 0.15 & 0.46\tablenotemark{aa} & 0.02\tablenotemark{a}  & 2.5 & isochrone & \citet[][]{Balona13} \\ 
  12.00$\pm$0.06\tablenotemark{aa} & 12.40$\pm$0.05 & 0.13$\pm$0.01 & 0.40$\pm$0.03\tablenotemark{aa} & 0.022$\pm$0.004\tablenotemark{a} & 1.9$\pm$0.1 & isochrone & \citet[][]{Wu13} \\ 
  \textbf{11.83$\pm$0.14} & \textbf{12.27$\pm$0.02} & \textbf{0.142$\pm$0.044} & \textbf{\nodata} & \textbf{$+$0.09$\pm$0.03\tablenotemark{b}} & \textbf{\nodata} & \textbf{asteroseismology} & \textbf{The present work\tablenotemark{bb}}\\
\textbf{11.88$\pm$0.14} & \textbf{12.32$\pm$0.03} & \textbf{0.142$\pm$0.044} & \textbf{\nodata} & \textbf{$+$0.09$\pm$0.03\tablenotemark{b}} & \textbf{\nodata} & \textbf{asteroseismology} & \textbf{The present work\tablenotemark{cc}}
\enddata
\tablenotetext{a}{Metal fraction $Z$.}
\tablenotetext{b}{Metallicity [Fe/H].}
\tablenotetext{c}{Solar metallicity, corresponding [Fe/H]=0.0.}
\tablenotetext{d}{Based on Yale isochrone models.}
\tablenotetext{e}{Based on VandenBerg isochrone models.}
\tablenotetext{aa}{Calculated with Equations \eqref{eq_dm-a} and/or \eqref{eq_extinction}.}
\tablenotetext{bb}{Based on classical relation (Equation \eqref{eq-log2}).}
\tablenotetext{cc}{Based on new relation (Equation \eqref{eq-log3}).}
\tablecomments{Column 1---True distance modulus (${\rm (m-M)_{0}}$); Column 2---Apparent distance modulus (${\rm (m-M)_{V}}$); Column 3---Interstellar reddening ($E(B-V)$); Column 4---Interstellar extinction ($A_{V}$); Column 5---Metallicity ($Z$ (metal fraction) or [Fe/H]); Column 6---Cluster ages; Column 7---Methods used to determine distance modulus; Column 8---Reference.}
\end{deluxetable*}
\renewcommand{\arraystretch}{1}

In the \textit{Kepler} field of view there are two open clusters NGC~6791 and NGC~6819 in which solar-like oscillations have been observed  for a number of red-giant stars \citep{Stello10,Stello11a,Stello11b,Hekker11b,basu11,Miglio12,Corsaro12,Balona13,Wu13}. An overview of earlier work regarding distance moduli, interstellar extinctions/reddenings, ages and metallicities presented in the literature for these clusters is provided in Table~\ref{table_liter}. In short: NGC~6791 is one of the oldest \citep[$6\sim8$ Gyr, e.g.][]{Harris_Canterna81,Wu13} clusters with super-solar metallicity \citep[${\rm[Fe/H]}\approx0.3\sim0.4$ dex, e.g.][]{Carraro06,Brogaard11,Wu13}, with a true distance modulus in the range $12.9 \sim 13.1$ mag \citep{basu11,Miglio12,Wu13}.
NGC~6819 is an intermediate-age cluster \citep[$1.6\sim2.5$ Gyr, e.g.][]{rv98,Kalirai01,basu11,Wu13} with near-solar or slightly super-solar metallicity \citep[e.g.][]{Bragaglia01,Hole09,Warren&Cole09,Wu13}. The true distance modulus of this cluster is of the order of $11.8 \sim 12.0$ \citep[e.g.][]{basu11,Jeffries13,Balona13,Wu13}.

In this paper, we propose a new method to estimate the cluster distance modulus from global oscillation parameters (\dnu\ and \numax) and V photometry of cluster members of NGC~6791 and NGC~6819. This method is based on a relation between the frequency of maximum oscillation power \numax, the large frequency separation \dnu, the apparent magnitude $V$, the metallicity $Z$, and the distance modulus ${\rm (m-M)_{0}}$.

\section{Derivation of distance modulus relations}\label{sec-theory}

For solar-like oscillations, there are two scaling relations with respect to large frequency separation \dnu\ and the frequency of maximum oscillation power \numax. They are
\begin{equation}\label{eq-dnu}
\Delta\nu=\sqrt{\frac{M/M_{\sun}}{(R/R_{\sun})^3}}\Delta\nu_{\sun}
\end{equation}
and
\begin{equation}\label{eq-numax}
\nu_{\rm{max}}=\frac{M/M_{\sun}}{(R/R_{\sun})^2\sqrt{T_{\rm{eff}}/T_{\rm{eff},\sun}}}\nu_{\rm{max},\sun},
\end{equation}
which are described by \citet{kjeldsen95}. In the above equations, $\Delta\nu_{\sun}=134.88~\mu$Hz, $\nu_{\rm{max},\sun}=3120~\mu$Hz, and $T_{\rm{eff},\sun}=5777$ K, which are taken from \citet{Kallinger10}. The two equations are usually used to determine stellar parameters, such as, the mass $M$, radius $R$, mean density $\bar{\rho}$, surface gravity $g$. For the two scaling relations, many detailed discussions have been presented; for example, \citet[][]{Bedding-kje03}, \citet[][]{Stello08}, \citet[][]{Kallinger10}, \citet[][]{White11}, \citet[][]{Miglio12}, \citet[][]{Mosser13}, and \citet[][]{hekker13}. For this work we have decided to not include any of the proposed corrections \citep{White11,Miglio12,Mosser13} as there is no consensus in the literature of the size of the correction for red giant branch stars \citep{hekker13} to which we apply the scalings in the present study.

Combining the two equations (Equations \eqref{eq-dnu} and \eqref{eq-numax}) and the relation among the stellar luminosity $L$, the effective temperature \teff, and the radius $R$:
\begin{equation}\label{eq-L}
\log\frac{L}{L_{\sun}} = 2\log\frac{R}{R_{\sun}}+4\log\frac{T_{\rm{eff}}}{T_{\rm{eff},\sun}},
\end{equation}
we can obtain a relation
\begin{equation}\label{eq-log-d-n}
  24\log\nu_{\rm{max}}=28\log\Delta\nu+10\log M-3\log L,
\end{equation}
where all the variables (large frequency separation \dnu, frequency of maximum oscillation power \numax, stellar mass $M$, and luminosity $L$) are in units of the corresponding solar values.\footnote{In the following derivations and analyses all variables (such as, the large frequency separation \dnu, the frequency of maximum oscillation power \numax, the stellar mass $M$, the effective temperature \teff, the luminosity $L$) are in units of the corresponding solar values except when units are explicitly shown. In other word, we ignore the unit of variables in the procedure of derivation and restitute their units in the final equations.}

The relation between the absolute bolometric magnitude ${\rm M_{b,\star}}$ and stellar luminosity $L$ can be expressed as:
$$
{\rm M_{b,\star}-M_{b,\sun}}=-2.5\log(\frac{L}{L_{\sun}}),
$$
where, ${\rm M_{b,\sun}}=+4.75$ is the solar absolute bolometric magnitude. From this relation it follows that:
\begin{equation}\label{eq-L1}
\log L=0.4(4.75-{\rm M_{b,\star}}).
\end{equation}
In general, it is very difficult to detect the stellar bolometric magnitude in observations. The stellar radiation can in fact only be measured in a few specific spectral bands, such as V band. Thus, in Equation \eqref{eq-L1} the absolute bolometric magnitude ${\rm M_{b,\star}}$ needs to be replaced by the absolute apparent magnitude ${\rm M_{V}}$. In order to obtain the stellar bolometric magnitude, one has to introduce a new physical parameter --- the bolometric correction $BC$ --- the difference between ${\rm M_{b,\star}}$ and ${\rm M_{V}}$:
\begin{equation}\label{eq_mb}
{\rm M_{b,\star}=M_{V}}+BC.
\end{equation}

Corresponding to the absolute apparent magnitude ${\rm M_{V}}$, the parameter that can be detected by an observer is the apparent magnitude $V$. Due to the interstellar medium between the star and the observer, the value of apparent magnitude $V$ will be larger than the intrinsic value that is unaffected by the interstellar medium. Such an intrinsic value is usually named as the true apparent magnitude and denoted by $V_{0}$. The difference between the true and apparent magnitudes is the interstellar extinction $A_{\rm V}$. It can be expressed as
\begin{equation}\label{eq-av}
A_{\rm V}=V-V_{0}.
\end{equation}

According to the definition of distance modulus, the difference between $V_{0}$ and ${\rm M_{V}}$ is called the true distance modulus ${\rm (m-M)_{0}}$ and the difference between $V$ and ${\rm M_{V}}$ is called the apparent distance modulus ${\rm (m-M)_{V}}$. Combined with Equation \eqref{eq-av}, the relationship among $V_{0}$, $V$, ${\rm M_{V}}$, $A_{\rm V}$, ${\rm (m-M)_{0}}$, and ${\rm (m-M)_{V}}$ can be therefore expressed as
\begin{equation}\label{eq_dm-a}
\begin{split}
{\rm (m-M)_{0}} & =V_{0}-{\rm M_{V}}=V-{\rm M_{V}}-A_{\rm V} \\ &={\rm (m-M)_{V}}-A_{\rm V}.
\end{split}
\end{equation}

Combining Equation \eqref{eq_mb} with Equation \eqref{eq_dm-a}, we obtain the following relation
\begin{equation}\label{eq-mb-2}
  {\rm M_{b,\star}}=V-A_{V}-{\rm (m-M)_{0}}+BC.
\end{equation}
Therefore, combining Equation \eqref{eq-L1} with Equation \eqref{eq-mb-2} results in the following equation:
\begin{equation}\label{eq-LogL}
  \log L=0.4[4.75-V+A_{\rm V}-BC+{\rm (m-M)_{0}}].
\end{equation}

Finally, combining Equations \eqref{eq-log-d-n} and \eqref{eq-LogL}, gives the following relation:
\begin{equation}\label{eq-log1}
\begin{split}
  24\log\nu_{\rm{max}}&=28\log\Delta\nu+1.2(V+BC)-5.7 \\ &+10\log M-1.2{\rm (m-M)_{0}}-1.2A_{\rm V}.
\end{split}
\end{equation}
Note that from Equation \eqref{eq-log1} we can estimate the distance modulus \dm\ by use of the observation parameters (\dnu, \numax, and $V$), if we know the stellar mass $M$, the bolometric correction $BC$ and the extinction $A_{\rm V}$.

We can use \dnu\ and \numax\ to eliminate the mass $M$ from Equation \eqref{eq-log1}. In this way Equation \eqref{eq-log1} can be rewritten as
\begin{equation}\label{eq-log2}
\begin{split}
6\log\nu_{\rm{max}}+&15\log T_{\rm eff}=12\log\Delta\nu-1.2(V+BC)\\ & +1.2{\rm (m-M)_{0}}+1.2A_{\rm V}+5.7.
\end{split}
\end{equation}

Alternatively, Equation \eqref{eq-log2} can be directly derived from Equations \eqref{eq-dnu}, \eqref{eq-numax}, \eqref{eq-L}, \eqref{eq-L1} and \eqref{eq-mb-2}. Equation \eqref{eq-log2} shows that we can determine the distance modulus \dm\ from \dnu, \numax, $V$, \teff, and $BC$ and/or analyze the relation between the distance modulus \dm\ and interstellar extinction $A_{\rm V}$.

In an alternative approach, we use a relation for red giant branch (RGB) stars based on the Hayashi relation ($\sqrt{T_{\rm eff}} \sim g^{p}R^{q}$) derived by \citet[][Equation (11)]{Wu13}. This relation describes the stellar effective temperature \teff\ as a function of the stellar radius $R$, the stellar mass $M$, and the metallicity $Z$ (metal fraction) as follows:
\begin{equation}\label{eq-T}
0.5\log T_{\rm{eff}}= a\log R+b\log M+c\log Z+d,
\end{equation}
where, $a=-0.049$, $b=0.051$, $c=-0.022$, and $d=-0.008$ and metal fraction $Z$ with the unit of $Z_{\sun}=0.02$, \textbf{which are taken from \cite{Wu13}}. Combining Equations \eqref{eq-dnu}, \eqref{eq-numax}, and \eqref{eq-T}, the stellar mass $M$ can be expressed as:
\begin{equation}\label{eq-m}
M^{a+3b-1}=10^{-3d}\Delta\nu^{4+2a}\nu_{\rm max}^{-3}Z^{-3c},
\end{equation}
which corresponds to Equation (16) of \citet[][]{Wu13}.

This can be used to obtain a relation among \numax, \dnu, $Z$, $V$, $BC$, $A_{\rm V}$, and ${\rm (m-M)_{0}}$, by substituting Equation \eqref{eq-m} into Equation \eqref{eq-log1} to eliminate the mass $M$:
\begin{equation}\label{eq-log3}
\begin{split}
9.&482\log\nu_{\rm{max}}=15.549\log\Delta\nu- 1.2(V+BC) \\ &+0.737\log Z+1.2{\rm (m-M)_{0}}+1.2A_{\rm V}+5.968.
\end{split}
\end{equation}

\begin{figure}
  \begin{center}
  \includegraphics[scale=0.5,angle=-90]{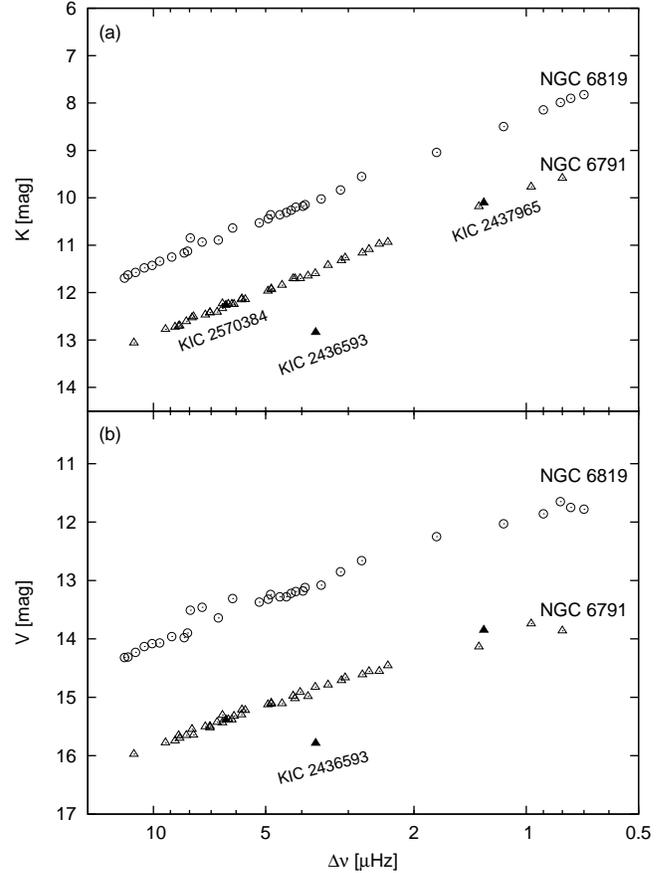}
  \caption{Similar to Figure 7 of \citet[][]{Stello11b}. Apparent magnitude vs. large frequency separation for NGC 6789 (open circle) and NGC 6791 (open triangle).}\label{fig-VK}
  \end{center}
\end{figure}

In fact, Equation \eqref{eq-log3} is not only applicable in the V band, but can also be used in other wavelength bands with a little modification. However in that case, $V$ and $A_{\rm V}$ should be replaced by $m_{\lambda}+(V-m_{\lambda})$ and $A_{\lambda}$, respectively, where $m_{\lambda}$ is the apparent magnitude of the $\lambda$ band, $(V-m_{\lambda})$ is the color excess between V and $\lambda$ band, and $A_{\lambda}$ is the interstellar extinction of the $\lambda$ band, or $V$, $BC$, and $A_{\rm V}$ should be replaced by $m_{\lambda}$, $BC_{\lambda}$, and $A_{\lambda}$, where $BC_{\lambda}$ is the general bolometric correction of $\lambda$ band, i.e., $BC_{\lambda}={\rm M_{b,\star}}-{\rm M}_{\lambda}$. As a result, Equation \eqref{eq-log3} accordingly becomes
\begin{equation}\label{eq-log4}
\begin{split}
&9.482\log\nu_{\rm{max}}=15.549\log\Delta\nu- 1.2[m_{\lambda}+(V-m_{\lambda}) \\ &+BC]+0.737\log Z+1.2{\rm (m-M)_{0}}+1.2A_{\lambda}+5.968
\end{split}
\end{equation}
or
\begin{equation}\label{eq-log5}
\begin{split}
9.&482\log\nu_{\rm{max}}=15.549\log\Delta\nu- 1.2(m_{\lambda}+BC_{\lambda}) \\ &+0.737\log Z+1.2{\rm (m-M)_{0}}+1.2A_{\lambda}+5.968.
\end{split}
\end{equation}
These equations (Equations \eqref{eq-log3}, \eqref{eq-log4}, and/or \eqref{eq-log5}) can be used to explain the trend between $V$ or $K$, and \dnu\ \citep[see Figure 7 in][and Figure \ref{fig-VK} in the present study]{Stello11b}.
In Figure 7 of \citet[][]{Stello11b}, these trends have similar slopes but different intercepts. From Equations \eqref{eq-log3}, \eqref{eq-log4}, and/or \eqref{eq-log5}, we suggest that those different intercepts of different clusters are due to those clusters having different metallicities $Z$, distance moduli ${\rm (m-M)_{0}}$, and interstellar extinctions $A_{\rm V}$.

Additionally, Equation \eqref{eq-log3} can be used to determine the cluster distance modulus ${\rm (m-M)_{0}}$ from \numax, \dnu, $V$, and cluster metallicity $Z$ (see Section~\ref{sec-nr}).

\section{Data Source and Cluster Parameters}\label{sec-data}

\begin{figure}
  \begin{center}
  \includegraphics[scale=0.5,angle=-90]{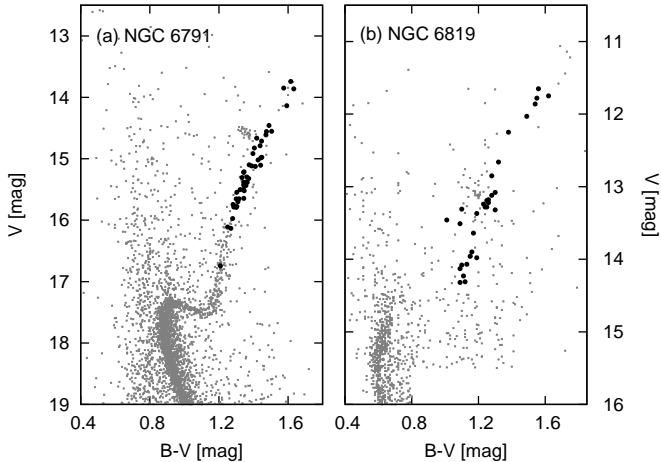}
  \caption{Color-Magnitude diagram (CMD). The photometric data in \textit{B} and \textit{V} band are derived from \citet{Stetson03} for NGC 6791 (panel (a)) and from \citet{Hole09} for NGC 6819 (panel (b)), respectively. The larger fulled points represent the investigated targets in the present study. }\label{fig-cmd}
  \end{center}
\end{figure}

In the present study, the investigated targets (see Figure \ref{fig-cmd}, larger fulled points) and their oscillation parameters are taken from \citet{Wu13} (for the detailed description with respect to the selection of targets and the analyses of observational data see \citet{Wu13}). The relative uncertainties of \dnu\ and \numax\ are about 1.2\% and 1.5\%, respectively.

To estimate the bolometric corrections $BC$ for the targets we use the $T_{\rm{eff}}:BC$ scales established by \citet[][hereafter F96]{Flower96} and the coefficients corrected and modified by \citet{Torres10}. At the same time, we use the color-temperature calibrations established by \citet[][hereafter RM05]{rm05} to estimate the effective temperatures. For the considered targets, the photometric data in B and V band are derived from \citet{Stetson03} for NGC~6791 and from \citet{Hole09} for NGC~6819 in the same way with \citet[][]{Wu13}. In addition, the K photometry is derived from the 2MASS catalog \citep{Skrutskie06}.  For the metallicity and the interstellar reddening, we adopt ${\rm[Fe/H]}=+0.29\pm0.10$ dex and $E(B-V)=0.16\pm0.025$ mag for NGC~6791. These values are obtained from spectroscopic observations \citep[][]{Brogaard11}. For NGC~681, we adopt ${\rm [Fe/H]}=+0.09\pm0.03$ dex and $E(B-V)=0.142\pm0.044$ mag. These values are obtained from high-dispersion spectroscopy of four clump stars \citep[][]{Bragaglia01}. The data sources and basic input parameters are listed in Table~\ref{table_1}. The reddening conversion $E(V-K)=2.72E(B-V)$ established by \citet{McCall04} is used.

Using the RM05 color-temperature relations ($(V-K):T_{\rm{eff}}$) we find that for the giants an uncertainty of 0.02 mag in $(V-K)$ and an uncertainty of 0.1 dex in [Fe/H] lead to an uncertainty of 15 K and 5 K in \teff, respectively. For the basic parameters of the clusters, we adopt 0.10 dex as the uncertainty of the metallicity [Fe/H], 0.02 mag as the uncertainties in $V$ and $K$ \citep[same as][]{Hekker11a}, and 0.04 mag as the uncertainty in the reddening $E(B-V)$. Combining this with the system uncertainty of 30 K (RM05), we obtain a total uncertainty in effective temperature: 5 K (metallicity) + 80 K (reddening) + 30 K ($(V-K)$) + 30 K (RM05) = 145 K.

Using the F96 $T_{\rm{eff}}:BC$ scales we find that for \teff\ ranging from 4000 K to 5500 K with an uncertainty of 145 K in \teff\ leads to an uncertainty of about 0.07 mag in the bolometric correction $BC$. There is no definite discussion with respect to the uncertainty in $T_{\rm{eff}}:BC$ scales. We adopt 0.05 mag as its system uncertainty from the results by F96 and the data of Table 2 of their paper. Therefore, we obtain a total uncertainty in the bolometric correction: 0.07 mag (\teff) + 0.05 mag (F96) = 0.12 mag.


The interstellar extinction ($A_{\rm V}$) is assumed to be the same for all stars in a cluster, because of the size of a cluster. It can be also expressed as a function of interstellar reddening $E(B-V)$:
\begin{equation}\label{eq_extinction}
A_{\rm V}=3.1E(B-V).
\end{equation}
More detailed discussions with respect to extinction and reddening have been presented by e.g., \citet{SM1979}, \citet{WD2001}, \citet{FM2003}, and \citet{Bilir2008}.

\renewcommand{\arraystretch}{1.1}
\begin{deluxetable}{lllll}
\tablecaption{Basic parameters and data sources of NGC 6791 and NGC 6819.\label{table_1} }
\tablewidth{0.48\textwidth}
\tablehead{
\colhead{Parameter} & \colhead{Value} & \colhead{Ref.} & \colhead{Value} & \colhead{Ref.}
}
\startdata
              &  NGC 6791            &     & NGC 6819               & \\
  \hline
  $V$         &  \nodata             & (1) & \nodata                & (2) \\
  $K$         &  \nodata             & (3) & \nodata                & (3) \\
  \numax      & \nodata              & (4) & \nodata                & (4) \\
  \dnu        & \nodata              & (4) & \nodata                & (4) \\
  $E(B-V)$    & $0.16\pm0.025$ mag   & (5) & $0.142\pm0.044$ mag    & (6) \\
${\rm[Fe/H]}$ & $0.29\pm0.10$ dex    & (5) & $0.09\pm0.03$ dex      & (6) 
\enddata
\tablerefs{
(1) \citet{Stetson03};
(2) \citet{Hole09};
(3) 2MASS;
(4) \citet[][]{Wu13};
(5) \citet[][]{Brogaard11};
(6) \citet[][]{Bragaglia01}.
}
\end{deluxetable}
\renewcommand{\arraystretch}{1}

\section{Distance moduli for NGC~6791 and NGC~6819}\label{sec-analysis}

In this section we refer to Equation \eqref{eq-log2} as the `classical relation' as this is based on the scaling relation by \citet{kjeldsen95}. Equation \eqref{eq-log3} is referred to as the `new relation' because it is based on the relations for red giant branch stars derived from the Hayashi track.

\subsection{Classical Relation}\label{sec-cr}

\renewcommand{\arraystretch}{1.1}
\begin{deluxetable}{lcr}
\tablewidth{0.48\textwidth}
\tablecaption{Fitting relations and fitting coefficients.\label{table_fit}}
\startdata
\hline \hline
     & \multicolumn{2}{l}{fit: $12\log\Delta\nu-1.2(V+BC)$} \\
 (1) & \multicolumn{2}{r}{$ =A_{1}[6\log\nu_{\mathrm{max}}+15\log T_{\rm eff}]+B_{1}$} \\
     & \multicolumn{2}{c}{underlying equation \eqref{eq-log2}} \\
       ~~ &  $A_{1}$                     &       $B_{1}$       \\
 \hline
 NGC~6791 & $1.004\pm0.015$      &     $-21.939\pm0.197$ \\
      ~   & 1.0\tablenotemark{a} &     $-21.995\pm0.034$ \\
 NGC~6819 & $0.992\pm0.008$      &     $-20.527\pm0.100$ \\
     ~    & 1.0\tablenotemark{a} &     $-20.429\pm0.024$ \\ 
\hline \hline

     & \multicolumn{2}{l}{fit: $15.549\log\Delta\nu-1.2(V+BC)$} \\
(2)  & \multicolumn{2}{r}{$=A_{2}[9.482\log\nu_{\mathrm{max}}]+B_{2}$} \\
     & \multicolumn{2}{c}{underlying equation \eqref{eq-log3}} \\
       ~~ &  $A_{2}$               &       $B_{2}$       \\
 \hline
 NGC~6791 & $1.014\pm0.009$        &     $-22.246\pm0.157$ \\
      ~   & 1.0\tablenotemark{a}   &     $-22.492\pm0.028$ \\
 NGC~6819 & $1.001\pm0.006$        &     $-20.804\pm0.108$ \\
     ~    & 1.0\tablenotemark{a}   &     $-20.820\pm0.026$
\enddata
\tablenotetext{a}{Predicted value.}
\tablecomments{The variables \numax, \dnu, and \teff\ are in solar units, while $V$ and $BC$ are expressed in magnitude.}
\end{deluxetable}
\renewcommand{\arraystretch}{1}

\begin{figure}
  \begin{center}
  \includegraphics[scale=0.58,angle=-90]{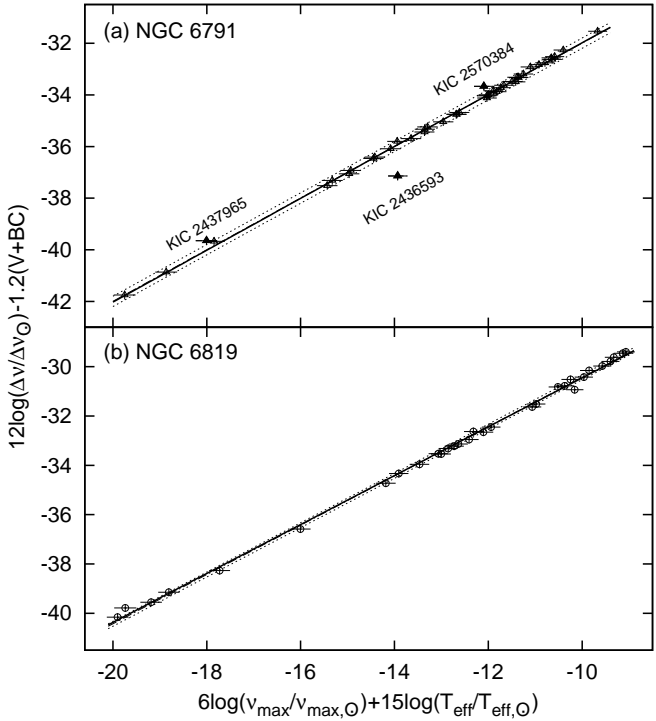}
  \caption{$12\log(\Delta\nu/\Delta\nu_{\sun})+1.2(4.75-V-BC)$ vs. $6\log(\nu_{\rm{max}}/\nu_{\rm{max,\sun}})+15\log(T_{\rm eff}/T_{\rm eff,\sun})$ for NGC~6791 (panel (a)) and NGC~6819 (panel (b)). The dash-dotted lines show the fits with fitting relation (1) of Table~\ref{table_fit}, the dashed lines show the corresponding $1\sigma$ uncertainties, and the solid lines show the fits of the theoretical prediction, i.e., the coefficient $A_{1}$ fixed to 1.0 and coefficient $B_{1}$ as a free parameter. Solid symbols indicated with KIC numbers are discussed in the text.}\label{fig1}
  \end{center}
\end{figure}

From Equation \eqref{eq-log2} it can be derived that stars in a cluster should show a linear relation when $12\log\Delta\nu-1.2(V+BC)$ is plotted as a function of $6\log\nu_{\rm{max}}+15\log T_{\rm eff}$, if they have the same distances and interstellar extinctions. This relation is shown in Figure \ref{fig1}. All investigated targets are indeed located on a linear relation which confirms the cluster memberships of the considered stars. Additionally, Equation \eqref{eq-log2} and the above analysis show that we can obtain the cluster distance modulus through fitting the relation $12\log\Delta\nu-1.2(V+BC)=A_{1}[6\log\nu_{\mathrm{max}}+15\log T_{\rm eff}]+B_{1}$ (fitting relation (1) of Table~\ref{table_fit}). Combining Equation \eqref{eq-log2} and the fitting relation, we obtain a relation between distance modulus \dm, interstellar extinction $A_{\rm V}$, and the fitting coefficient $B_{1}$:
\begin{equation}\label{eq-DM-B1}
B_{1}=-1.2{\rm (m-M)_{0}}-1.2A_{\rm V}-5.7.
\end{equation}
For the observations, the fitting coefficient $A_{1}$ is $1.004\pm0.015$ for NGC~6791 and $0.992\pm0.008$ for NGC~6819, respectively. These are in good agreement with the theoretical prediction 1.0 (see Figure \ref{fig1} and fitting (1) of Table~\ref{table_fit}). In Figure \ref{fig1}, the two fitted lines --- the fit with both $A_{1}$ and $B_{1}$ as free parameters and the fit with $A_{1}$ fixed to 1.0 --- are in good agreement with each other and with the data. Therefore, combining the value of fitted coefficients $B_{1}$, the interstellar extinction $A_{\rm V}$, and Equation \eqref{eq-DM-B1}, we can obtain the cluster true distance modulus \dm.

For NGC 6791, combining Equations \eqref{eq-DM-B1} with \eqref{eq_extinction} and substituting the values of $B_{1}$ and $E(B-V)$, we obtain a cluster true distance modulus ${\rm (m-M)_{\rm 0,6791}}=13.08\pm0.08$ mag and the corresponding apparent distance modulus of  ${\rm (m-M)_{\rm V,6791}}=13.58\pm0.03$ mag. For NGC 6819, we obtain the cluster true distance modulus ${\rm (m-M)_{\rm 0,6819}}=11.83\pm0.14$ mag and the apparent distance modulus ${\rm (m-M)_{\rm V,6819}}=12.27\pm0.02$ mag in the same way. These results are listed in Table~\ref{table_liter} in bold font.

It can be found from the panel (a) of Figure \ref{fig1} that for NGC~6791 there are three data points (KIC 2436593, KIC 2437965 and KIC 2570384) deviating from the fits. They are denoted by filled points in Figure \ref{fig1}. Those deviations may be caused by the fact that these targets are potential blends \citep[see][for detailed discussion]{Stello11b}. Besides the blending, the interstellar extinction may be another factor for those deviations.

In the above analysis we do not calculate individual stellar distance moduli of the cluster stars, but regard all considered targets in a cluster as an entity. This is a novel way to calculate the average distance modulus of a cluster. It fully reflects the characteristic of member stars in a cluster.

In the above analysis, all results are based on Equation \eqref{eq-log2}, which is derived from the solar-like oscillations and photometric observations. It is therefore suitable for stars showing solar-like oscillations. However, uncertainties may vary as a function of the accuracy of the scaling relations \citep{White11,Miglio12,Mosser13,hekker13}

\subsubsection{Sources of Uncertainties in the Classical Relation}\label{sec-uncr}

\begin{figure}
  \begin{center}
  \includegraphics[scale=0.42,angle=-90]{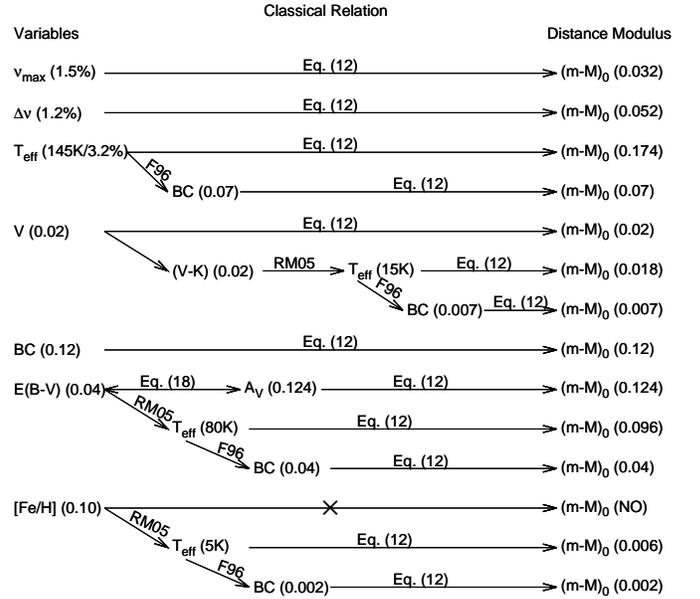}
  \caption{The sources and propagation of uncertainties for the classical relation (Equation \eqref{eq-log2}). Uncertainties propagate from left to right directly (one line connects the variables and the distance modulus directly) or indirectly. `$\times$' denotes that the distance modulus is not directly affected by the variable. Note that in the uncertainty analyses, we use a characteristic temperature $\bar{T}_{\rm eff}\approx4500$ K as a characteristic stellar effective temperature.}\label{fig-tree}
  \end{center}
\end{figure}

Distance moduli obtained with the classical relation are mainly affected by the uncertainties in $BC$, \teff, and $E(B-V)$ (see Figure \ref{fig-tree}). Fundamentally, the major uncertainty comes from the uncertainty of $E(B-V)$, because $E(B-V)$ significantly affects the effective temperature \teff\ and further affects the bolometric correction $BC$. For example, a change of 0.01 mag in $E(B-V)$ directly leads to a change of 0.03 mag in \dm\ through Equations \eqref{eq-DM-B1} and/or \eqref{eq_extinction} and to a change of about 20 K in \teff\ through the RM05 color-temperature relation accordingly. A change of 20 K in \teff\ will directly lead to a change of about 0.03 mag in \dm\ through Equation \eqref{eq-log2} and to a change of about 0.01 mag in $BC$ through F96 \teff\ :$BC$ scale. Furthermore, a change of 0.01 mag in $BC$ will lead to a change of 0.01 mag in \dm. Summarizing: a change of 0.01 mag in $E(B-V)$ will lead to a change of at least 0.04 mag in \dm. The influence of $E(B-V)$ for our results is complicated as well as significant and cannot be ignored.

Compared to the influence of other uncertainties, the influence of metallicity can be ignored, because it only slightly affects the effective temperature \teff\ and bolometric correction $BC$. For example, a change of 0.1 dex in [Fe/H] will lead to a change of 5 K in \teff. The change of 5 K will lead to a change of about 0.0023 mag in $BC$. The change of 0.1 dex in [Fe/H] will therefore lead to a change of about 0.0064 mag in ${\rm (m-M)_{0}}$ in total.

\subsection{New Relation}\label{sec-nr}

In Equation \eqref{eq-log3}, both the metallicity $Z$ and distance modulus \dm\ are assumed to be constant for a cluster. Here, we do not consider possibilities of stellar regeneration in a cluster and mergers between two or more clusters. As a result, we do not need to calculate these parameters individually for member stars of a cluster. In addition, the metallicity $Z$ can be obtained from spectroscopic observations. Therefore, using Equation \eqref{eq-log3} to determine the cluster distance modulus can be a convenient and effective method.

\begin{figure}
  \begin{center}
  \includegraphics[scale=0.58,angle=-90]{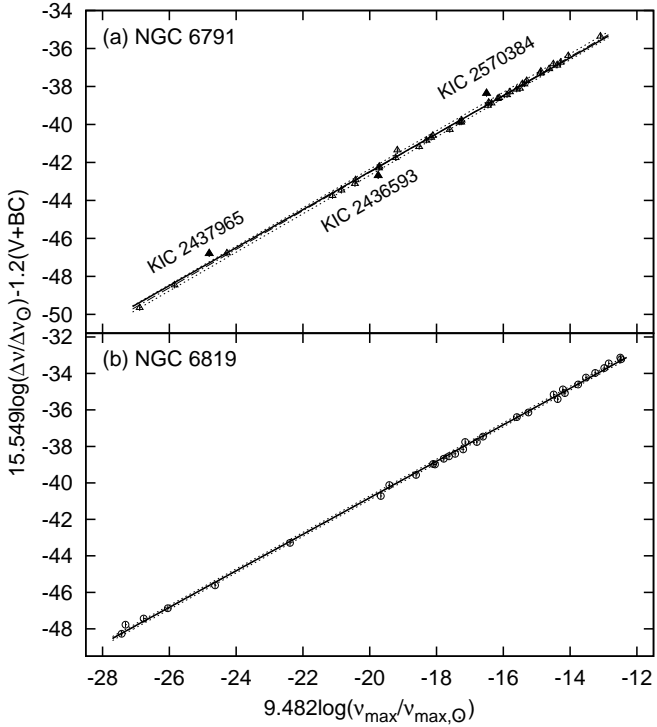}
  \caption{Similar to Fig.~\ref{fig1}, but now with $15.549\log(\Delta\nu/\Delta\nu_{\sun})-1.2(V+BC)$ vs. $9.482\log(\nu_{\rm{max}}/\nu_{\rm{max,\sun}})$ and the fits corresponding to relation (2) of Table~\ref{table_fit}. In addition, the filled symbols represent outliers (see text for more details).}\label{fig2}
  \end{center}
\end{figure}

From Equation \eqref{eq-log3}) it follows that stars should follow a linear relation when $15.549\log\Delta\nu-1.2(V+BC)$ is plotted as a function of $9.482\log\nu_{\rm{max}}$, if they have the same distances, metallicities and interstellar extinctions. This is indeed shown in Figure \ref{fig2} confirming the cluster membership of the considered stars. Additionally, we can estimate the cluster distance modulus or metallicity through fitting the relation $15.549\log\Delta\nu-1.2(V+BC)=A_{2}[9.482\log\nu_{\rm{max}}]+B_{2}$ (fitting relation (2) of Table~\ref{table_fit}). Combining Equation \eqref{eq-log3} and the fitting relation, we obtain a relation with respect to the true distance modulus \dm, metallicity $Z$, interstellar extinction $A_{\rm V}$, and the fitting coefficient $B_{2}$:
\begin{equation}\label{eq-dm-B2}
B_{2}=-0.737\log Z-1.2{\rm (m-M)_{0}}-1.2A_{\rm V}-5.968.
\end{equation}
From the analysis of the observational data with fitting relation (2) of Table~\ref{table_fit}, and substituting the corresponding coefficients into Equation \eqref{eq-dm-B2}, the cluster distance modulus can be obtained.

For NGC~6791, we obtain the fitted coefficient $A_{2}$ to be $1.014\pm0.009$, which is within $2\sigma$ of the theoretically predicted value of 1.0 (see panel (a) of Figure \ref{fig2} and fitting (2) of Table~\ref{table_fit}). The consistency between the fits and the data allow us to combine the fitted coefficient $B_{2}$, Equation \eqref{eq-dm-B2}, and spectroscopic metallicity ([Fe/H]\footnote{In the present study, we adopt the relation ${\rm [Fe/H]}\approx \log(Z/Z_{\sun})$ to make transformation between [Fe/H] and metal fraction $Z$ approximatively.}) to obtain the cluster distance modulus. Here we use [Fe/H] \citep[$+0.29\pm0.10$ dex,][spectroscopy]{Brogaard11}. In this way we obtain the cluster true distance modulus ${\rm (m-M)_{0,6791}}=13.09\pm0.10$ mag and the corresponding apparent distance modulus ${\rm (m-M)_{V,6791}}=13.59\pm0.06$ mag. These results are listed in Table~\ref{table_liter} in bold font. It can be found from Table~\ref{table_liter} that the results obtained from the new relation (Equation \eqref{eq-log3}) are consistent with those previously obtained from classical relation (Equation \eqref{eq-log2}) and with the results from the literature.

We note that KIC 2436593, KIC 2437965, and KIC 2570384 are also discrepant in this analysis, as was the case in the analysis using the classical relation.


For NGC~6819, the fitted coefficient $A_{2}$ is $1.001\pm0.006$, which is in good agreement with the theoretically predicted value of 1.0 (see panel (b) of Figure \ref{fig2} and fitting (2) of Table~\ref{table_fit}). Substituting the metallicity $\rm{[Fe/H]}=+0.09\pm0.03$ dex \citep[][high-dispersion spectroscopy]{Bragaglia01} and the corresponding value of fitted coefficient $B_{2}$ into Equation \eqref{eq-dm-B2} and combining Equation \eqref{eq_extinction} and the value of corresponding reddening $E(B-V)$, we obtain the cluster true distance modulus ${\rm (m-M)_{0,6819}} =11.88\pm0.14$ mag and its corresponding apparent distance modulus ${\rm (m-M)_{V,6819}} =12.32\pm0.03$ mag. They are listed in Table~\ref{table_liter} in bold font. It can be noted from Table~\ref{table_liter} that these values are consistent with results from the classical relation and with results from the literature.


\subsubsection{Influence of metallicity {\rm[Fe/H]} on the New Relation}\label{sec-influence}

\begin{figure}
  \begin{center}
  \includegraphics[scale=0.42,angle=-90]{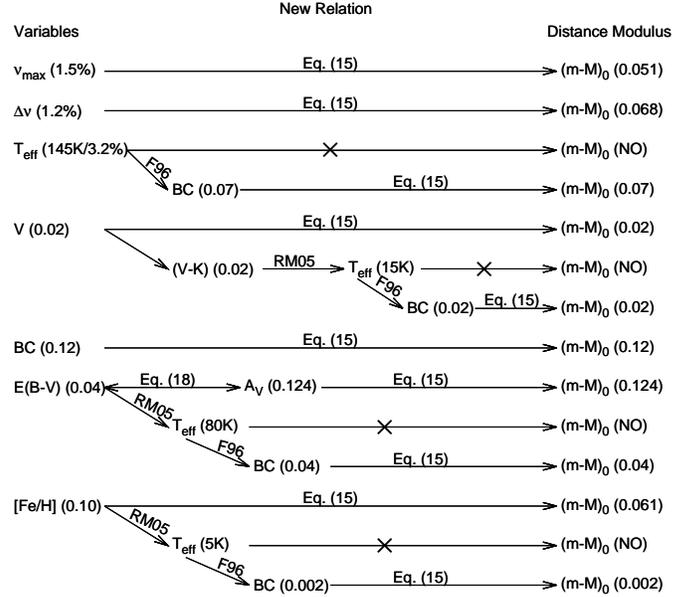}
  \caption{Similar to Figure \ref{fig-tree}, but for the new relation (Equation \eqref{eq-log3}).}\label{fig-tree1}
  \end{center}
\end{figure}

Equations \eqref{eq-log3} and \eqref{eq-dm-B2} show that the results determined from the new relation are mainly affected by two factors --- metallicity [Fe/H] and reddening $E(B-V)$. For reddening $E(B-V)$, the influence on the results is similar as its influence on the classical relation (Equation \eqref{eq-log2}), except for the effective temperature as this is not a parameter in the new relation (for detailed analyses of the sources and propagation of uncertainty, see Figure \ref{fig-tree1}).

In order to determine cluster distance moduli, the metallicity [Fe/H] appears two times in the analysis of the new relation: firstly, the RM05 color--temperature calibration is dependent on metallicity and on the stellar bolometric correction $BC$; secondly, [Fe/H] is a key parameter in the determination of distance modulus using Equation \eqref{eq-dm-B2}. An uncertainty of 0.1 dex in the metallicity [Fe/H] leads to an uncertainty of about 5 K in \teff\  from the RM05 color--temperature calibration (see Section \ref{sec-data} and Figure \ref{fig-tree1}). The change of 5 K in \teff\ further leads to a change of less then 0.003 mag in the bolometric correction $BC$. Such a small change can be ignored compared to the obtained uncertainty of 0.12 mag in $BC$. In Equation \eqref{eq-dm-B2}, the change of 0.1 dex in [Fe/H] will lead to a change of about 0.06 mag in distance modulus. In other words, a change of 0.1 dex in [Fe/H] only leads to about 0.06 mag in distance modulus for the method of the present study. Comparing the change of less than 0.003 mag with the change of 0.06 mag in distance modulus due to the change of 0.1 dex in [Fe/H], we may conclude that the influence of metallicity in the new relation is dominated by the term $0.737\log(Z)$ in Equation \eqref{eq-log3} and \eqref{eq-dm-B2}. \textbf{Table~\ref{table_compare} is given to represent the influence of [Fe/H] on distance modulus of NGC 6791 in new relation. For NGC 6819, such influences are similar with NGC 6791.}


\renewcommand{\arraystretch}{1.1}
\begin{deluxetable}{lcr}
\tablewidth{0.47\textwidth}
\tablecaption{The Influence of ${\rm[Fe/H]}$ on Distance Moduli in new relation for Cluster NGC~6791.\label{table_compare}}
\tablehead{
\colhead{[Fe/H]\tablenotemark{a}} & \colhead{${\rm (m-M)_{V}}$\tablenotemark{b}}& \colhead{${\rm (m-M)_{0}}$\tablenotemark{b}} \\
\colhead{[dex]}  & \colhead{[mag]} & \colhead{[mag]} }
\startdata
$+$0.39$\pm$0.05 & 13.53$\pm$0.04 & 13.03$\pm$0.09 \\
$+$0.35$\pm$0.02 & 13.56$\pm$0.03 & 13.06$\pm$0.08 \\
$+$0.30$\pm$0.08 & 13.59$\pm$0.05 & 13.09$\pm$0.10 \\
\textbf{$+$0.29$\pm$0.10\tablenotemark{c}} & \textbf{13.59$\pm$0.06} & \textbf{13.09$\pm$0.10}
\enddata
\tablenotetext{a}{Input parameter.}
\tablenotetext{b}{Output parameter.}
\tablenotetext{c}{Adopted in the present study.}
\end{deluxetable}
\renewcommand{\arraystretch}{1}

It can be noted from the current analyses (Table~\ref{table_compare}) that the distance moduli are only slightly affected by the metallicity [Fe/H] in our new method. This is because the stellar bolometric correction $BC$ in Equation \eqref{eq-log3} is only slightly dependent on the metallicity [Fe/H], and the weight of \dm\ in Equation \eqref{eq-dm-B2} is about two times that of [Fe/H]. We can therefore obtain a more precise result for the cluster distance moduli by use of this method.



From the above analyses it has been shown that our new method is self-consistent in constraining the clusters distance moduli and their metallicities. It can therefore also be used to estimate the cluster metallicity [Fe/H].

\section{Discussions}\label{sec-discussions}

In the present study, we use two different relations --- the so-called `classical relation' (Equation \eqref{eq-log2}) and `new relation' (Equation \eqref{eq-log3}) --- to determine the distance moduli of cluster NGC~6791 and NGC~6819, respectively. The former relation (classical relation) is derived from solar-like oscillations and photometric observations and the latter relation (new relation) is derived from solar-like oscillations, photometric observations, and the theory of stellar structure and evolution of red giant branch stars. Thus, the former relation is in theory applicable to all stars with solar-like oscillations, while the latter relation is only available to the red giant branch stars. In the analysis of these two relations, we always regard all considered stars in a cluster as an entity. This is a novel way to deal with the cluster members and to calculate the average distance modulus of a cluster. It fully reflects the characteristic of member stars in a cluster.

Equation \eqref{eq-log3} can on the one hand be used to interpret the correlation between the apparent magnitude and the large frequency separations in Figure \ref{fig-VK} \citep[also see Figure 7 of][]{Stello11b}. In Figure \ref{fig-VK}, the V band clearly shows larger scatter than K band. \citet[][]{Stello11b} suggest that this phenomenon is due to the fact that the V band has stronger sensitivity to differential interstellar reddening. Indeed, Equations \eqref{eq-log4} or \eqref{eq-log5} give support for this point. Besides this, the blending may be another factor that may change the apparent magnitude and the color excess between different bands. 
On the other hand, we use it to determine clusters distance moduli.

We have attempted to analyze the sources and effects of uncertainties. We do not analyze the observational uncertainties of \numax, \dnu, and $V$, since they affect the results in a similar way in the classical and new relations. Additionally, we do not take into account uncertainties in stellar radii originating from different definitions. Strictly speaking, the effective radius $R$ is different from the asteroseismic radius $R_{\rm seismic}$. In the present study, to derive the new relation and to determine the cluster distance moduli we assume that they are equal (in Equations \eqref{eq-dnu} and \eqref{eq-numax} and Equations \eqref{eq-L} and \eqref{eq-T}). Therefore, the difference between the two different radii may lead to small systemic uncertainties \citep[Beno\^{\i}t Mosser 2014; private communication][]{} in the new relation. 

For the two relations (Equations \eqref{eq-log2} and \eqref{eq-log3}), it can be noted that the classical relation is significantly affected by the uncertainty in effective temperature \teff. In the new relation, however, this disappears and is replaced by \numax, \dnu, and $Z$. The values of \numax\ and \dnu\ have small uncertainties and $Z$ has a small weight compared to the other variables. Hence, the distance modulus only slightly depends on the metallicity in the new relation.

In Figure \ref{fig-VK}, Figure \ref{fig1} (panel a), and Figure \ref{fig2} (panel a) there are three stars, which deviate from the linear relation are indicated with filled dots. These stars are KIC 2436593, KIC 2437965, and KIC 2570384 of NGC 6791. 
These three targets are potential blends \citet[][]{Stello11b}. Additionally, in Section \ref{sec-cr} we suggested that the interstellar extinction may be another factor contributing to the deviations. However, these two factors are not sufficient to interpret those deviations in the three figures (Figures \ref{fig-VK}, \ref{fig1}, and \ref{fig2}) simultaneously. 
One additional contribution can be due to different evolutionary processes that have taken place in these stars.

\section{Summary and Conclusions}\label{sec_summary}

From the global oscillation parameters (large frequency separation \dnu\ and frequency of maximum oscillation power \numax) and photometry data (apparent magnitude \textit{V}), we have determined the distance moduli for clusters NGC~6791 and NGC~6819, applying a new method, which regards all stars in a cluster as one entity and determine a mean value of the distance modulus but do not calculated individual distance moduli for the stars. This fully reflects the characteristic of member stars in a cluster as a natural sample. From this investigation we conclude the following:

\rmnum{1}: Based on the solar-like oscillations and photometric observations, we have derived relation $6\log\nu_{\rm{max}}+15\log T_{\rm eff}=12\log\Delta\nu+1.2(4.75-V-BC)+1.2{\rm (m-M)_{0}}+1.2A_{\rm V}$. We then verified this relation using observational data, and determined the cluster distance moduli of NGC~6791 and NGC~6819.

\rmnum{2}: Based on the solar-like oscillations, photometric observations, and the theory of stellar structure and evolution of red giant stars, we have obtained a new relation $9.482\log\nu_{\rm{max}}=15.549\log\Delta\nu- 1.2(V+BC)+0.737\log Z+1.2{\rm (m-M)_{0}}+1.2A_{\rm V}+5.968$. We have verified this relation using observational data.

\rmnum{3}: Based on the new relations, we have interpreted the trends between the apparent magnitude and larger frequency separation. At the same time, we have determined the cluster apparent distance moduli to be $13.59\pm0.06$ mag for NGC~6791 and $12.32\pm0.03$ mag for NGC~6819, respectively. Accordingly the corresponding true distance modulus is $13.09\pm0.10$ mag for NGC~6791 and $11.88\pm0.14$ mag for NGC~6819, respectively.

\rmnum{4}: We have found that the influence of $E(B-V)$ for the distance modulus is very complicated and can not be neglected for the classical relation. The change of 0.01 mag in $E(B-V)$ will lead to an uncertainty of at least  0.04 mag in \dm. The contribution of \teff\ to the uncertainty is considerable in the classical relation, while it is not present in the new relation. Additionally, we have found that the distance modulus only slightly depends on the metallicity in the new relation.

\rmnum{5}: The new method presented here could be used as a discrimination tool to determine the membership of cluster stars in the same way as the asteroseismic method of \citet{Stello11b}.

\acknowledgments
This work is co-sponsored by the NSFC of China (Grant Nos. 11333006 and 10973035), and by the Chinese Academy of Sciences (Grant No. KJCX2-YW-T24). The authors express their sincere thanks to NASA and the \kepler\ team for allowing them to work with and analyze the \kepler\ data making this work possible. The \keplermission\ is funded by NASA's Science Mission Directorate. The authors also express their sincere thanks to Prof. Achim Weiss and Beno\^{\i}t Mosser for their instructive advice and productive suggestions. In addition, fruitful discussions with J. Su are highly appreciated. SH acknowledges support from the European Research Council under the European Community's Seventh Framewrok Programme (FP7/2007-2013) / ERC grant agreement no 338251 (StellarAges) and  from Deutsche Forschungsgemeinschaft (DFG) under grant SFB 963/1 ``Astrophysical flow instabilities and turbulence''. The authors are grateful to the anonymous referee for useful comments that significantly improved the paper.


\begin{thebibliography}{}


\bibitem[{{Anthony-Twarog}(1984){Anthony-Twarog,B.~J.}}]{Anthony-Twarog84}
 {Anthony-Twarog, B.~J.} 1984, \baas, 16, 504


\bibitem[{{Anthony-Twarog \& Twarog}(1985){Anthony-Twarog, B.~J., \& Twarog, B.~A.}}]{Anthony-Twarog_Twarog85}
     {Anthony-Twarog, B.~J., \& Twarog, B.~A.} 1985, \apj, 291, 595

\bibitem[{{Anthony-Twarog et al.}(2007){Anthony-Twarog, B.~J., Twarog, B.~A., \& Mayer, L.}}]{Anthony-Twarog07}
     {Anthony-Twarog, B.~J., Twarog, B.~A., \& Mayer, L.} 2007, \aj, 133, 1585

\bibitem[{{Anthony-Twarog et al.}(2013){Anthony-Twarog, B.~J., Deliyannis, C.~P., Rich, E., \& Twarog, B.~A.}}]{Anthony-Twarog13}
     {Anthony-Twarog, B.~J., Deliyannis, C.~P., Rich, E., \& Twarog, B.~A.} 2013, \apjl, 767, L19


\bibitem[{{Auner}(1974){Auner, G.}}]{Auner74}
 {Auner, G.} 1974, \aaps, 13, 143

\bibitem[{{Baglin et al.}(2006){Baglin, A., Auvergne, M., Barge, P., Deleuil, M., Catala, C., Michel, E., Weiss, W., \& COROT Team}}]{Baglin06}
 {Baglin, A., Auvergne,M., Barge, P., et al.} 2006, in Proc. of The CoRoTMission Pre-Launch Status¡ªStellar Seismology and Planet Finding, ed.M. Fridlund, A. Baglin, J. Lochard, \& L. Conroy (ESA-SP 1306; Noordwijk: ESA), 33

\bibitem[{{Balona et al.}(2013){Balona, L.~A., Medupe, T., Abedigamba, O.~P., Ayane, G., Keeley, L., Matsididi, M., Mekonnen, G., Nhlapo, M.~D., \& Sithole, N.}}]{Balona13}
     {Balona, L.~A., Medupe, T., Abedigamba, O.~P., et al.} 2013, \mnras, 430, 3472

\bibitem[{{Basu et al.}(2011){Basu, S., Grundahl, F., Stello, D., Kallinger, T., Hekker, S., Mosser, B., Garc{\'{\i}}a, R.~A., Mathur, S., Brogaard, K., Bruntt, H., Chaplin, W.~J., Gai, N., Elsworth, Y., Esch, L., Ballot, J., Bedding, T.~R., Gruberbauer, M., Huber, D., Miglio, A., Yildiz, M., Kjeldsen, H., Christensen-Dalsgaard, J., Gilliland, R.~L., Fanelli, M.~M., Ibrahim, K.~A., \& Smith, J.~C.}}]{basu11}
 {Basu, S., Grundahl, F., Stello, D., et al.} 2011, \apjl, 729, L10

\bibitem[{{Bedding \& Kjeldsen}(2003){Bedding, T.~R., \& Kjeldsen, H.}}]{Bedding-kje03}
 {Bedding, T.~R., \& Kjeldsen, H.} 2003, PASA, 20, 203

\bibitem[{{Bedin et al.}(2005){Bedin, L.~R., Salaris, M., Piotto, G., King, I.~R., Anderson, J., Cassisi, S., \& Momany, Y.}}]{Bedin05}
     {Bedin, L.~R., Salaris, M., Piotto, G., et al.} 2005, \apjl, 624, L45

\bibitem[{{Bedin et al.}(2008){Bedin, L.~R., King, I.~R., Anderson, J., Piotto, G., Salaris, M., Cassisi, S., \& Serenelli, A.}}]{Bedin08}
     {Bedin, L.~R., King, I.~R., Anderson, J., et al.} 2008, \apj, 678, 1279




\bibitem[{{Bilir et al.}(2008){Bilir, S., Ak, S., Karaali, S., Cabrera-Lavers, A., Chonis, T.~S., \& Gaskell, C.~M.}}]{Bilir2008}
     {Bilir, S., Ak, S., Karaali, S., et al.} 2008, \mnras, 384, 1178



\bibitem[{{Bragaglia et al.}(2001){Bragaglia, A., Carretta, E., Gratton, R.~G., Tosi, M., Bonanno, G., Bruno, P., Cal{\`i}, A., Claudi, R., Cosentino, R., Desidera, S., Farisato, G., Rebeschini, M., \& Scuderi, S.}}]{Bragaglia01}
 {Bragaglia, A., Carretta, E., Gratton, R.~G., et al.} 2001, \aj, 121, 327

\bibitem[{{Brogaard et al.}(2011){Brogaard, K., Bruntt, H., Grundahl, F., Clausen, J.~V., Frandsen, S., Vandenberg, D.~A., \& Bedin, L.~R.}}]{Brogaard11}
     {Brogaard, K., et al.} 2011, \aap, 525, A2




\bibitem[{{Burkhead}(1971){Burkhead, M.~S.}}]{Burkhead71}
 {Burkhead, M.~S.} 1971, \aj, 76, 251


\bibitem[{{Buzasi} {et~al.}(2000){Buzasi}, {Catanzarite}, {Conrow}, {et~al.}}]{Buzasi00}
{Buzasi}, D. L., {Catanzarite}, J., {Conrow}, T., {et~al.} 2000, \apj, 532, L133



\bibitem[{{Carney et al.}(2005){Carney, B.~W., Lee, J.-W., \& Dodson, B.}}]{Carney05}
     {Carney, B.~W., Lee, J.-W., \& Dodson, B.} 2005, \aj, 129, 656

\bibitem[{{Carraro et al.}(2006){Carraro, G., Villanova, S., Demarque, P., McSwain, M.~V., Piotto, G., \& Bedin, L.~R.}}]{Carraro06}
     {Carraro, G., Villanova, S., Demarque, P., et al.} 2006, \apj, 643, 1151

\bibitem[{{Chaboyer et al.}(1999){Chaboyer, B., Green, E.~M., \& Liebert, J.}}]{Chaboyer99}
     {Chaboyer, B., Green, E.~M., \& Liebert, J.} 1999, \aj, 117, 1360

\bibitem[{{Chaplin et al.}(2011){Chaplin, W.~J., Kjeldsen, H., Christensen-Dalsgaard, J., Basu, S., Miglio, A., Appourchaux, T., Bedding, T.~R., Elsworth, Y., Garc{\'{\i}}a, R.~A., Gilliland, R.~L., Girardi, L., Houdek, G., Karoff, C., Kawaler, S.~D., Metcalfe, T.~S., Molenda-{\.Z}akowicz, J., Monteiro, M.~J.~P.~F.~G., Thompson, M.~J., Verner, G.~A., Ballot, J., Bonanno, A., Brand{\~a}o, I.~M., Broomhall, A.-M., Bruntt, H., Campante, T.~L., Corsaro, E., Creevey, O.~L., Do{\u g}an, G., Esch, L., Gai, N., Gaulme, P., Hale, S.~J., Handberg, R., Hekker, S., Huber, D., Jim{\'e}nez, A., Mathur, S., Mazumdar, A., Mosser, B., New, R., Pinsonneault, M.~H., Pricopi, D., Quirion, P.-O., R{\'e}gulo, C., Salabert, D., Serenelli, A.~M., Silva Aguirre, V., Sousa, S.~G., Stello, D., Stevens, I.~R., Suran, M.~D., Uytterhoeven, K., White, T.~R., Borucki, W.~J., Brown, T.~M., Jenkins, J.~M., Kinemuchi, K., Van Cleve, J., \& Klaus, T.~C.}}]{Chaplin11}
 {Chaplin, W.~J., Kjeldsen, H., Christensen-Dalsgaard, J., et al.} 2011, Science, 332, 213









\bibitem[{{Corsaro} {et~al.}(2012){Corsaro}, {Stello}, {Huber}, {Bedding},  {Bonanno}, {Brogaard}, {Kallinger}, {Benomar}, {White}, {Mosser}, {Basu},  {Chaplin}, {Christensen-Dalsgaard}, {Elsworth}, {Garc{\'{\i}}a}, {Hekker},
  {Kjeldsen}, {Mathur}, {Meibom}, {Hall}, {Ibrahim}, \& {Klaus}}]{Corsaro12}
{Corsaro}, E., {Stello}, D., {Huber}, D., {et~al.} 2012, \apj, 757, 190


\bibitem[{{de Marchi et al.}(2007){de Marchi, F., Poretti, E., Montalto, M., Piotto, G., Desidera, S., Bedin, L.~R., Claudi, R., Arellano Ferro, A., Bruntt, H.,
\& Stetson, P.~B.}}]{de-Marchi07}
 {de Marchi, F., Poretti, E., Montalto, M., et al.} 2007, \aap, 471, 515


\bibitem[{{Fiorucci \& Munari}(2003){Fiorucci, M., \& Munari, U.}}]{FM2003}
 {Fiorucci, M., \& Munari, U.} 2003, \aap, 401, 781

\bibitem[{{Flower}(1996){Flower, P.~J.}}]{Flower96}
 {Flower, P.~J.} 1996, \apj, 469, 355 (F96)





\bibitem[{{Gao \& Chen}(2012){Gao, X.-H., \& Chen, L.}}]{Gao12}
 {Gao, X.-H., \& Chen, L.} 2012, Chinese Astronomy and Astrophysics, 36, 1

\bibitem[{{Garnavich et al.}(1994){Garnavich, P.~M., Vandenberg, D.~A., Zurek, D.~R., \& Hesser, J.~E.}}]{Garnavich94}
     {Garnavich, P.~M., Vandenberg, D.~A., Zurek, D.~R., \& Hesser, J.~E.} 1994, \aj, 107, 1097


\bibitem[{{Gilliland} {et~al.}(2010){Gilliland, R.~L., Brown, T.~M.,Christensen-Dalsgaard, J., Kjeldsen, H., Aerts, C., Appourchaux, T., Basu, S., Bedding, T.~R., Chaplin, W.~J., Cunha, M.~S., De Cat, P., De Ridder, J., Guzik, J.~A., Handler, G., Kawaler, S., Kiss, L., Kolenberg, K., Kurtz, D.~W., Metcalfe, T.~S., Monteiro, M.~J.~P.~F.~G., Szab{\'o}, R., Arentoft, T., Balona, L., Debosscher, J., Elsworth, Y.~P., Quirion, P.-O., Stello, D., Su{\'a}rez, J.~C., Borucki, W.~J., Jenkins, J.~M., Koch, D., Kondo, Y., Latham, D.~W., Rowe, J.~F., \& Steffen, J.~H.}}]{Gilliland10}
 {Gilliland, R.~L., Brown, T.~M.,Christensen-Dalsgaard, J., et~al.} 2010, \pasp, 122, 131




\bibitem[{{Grundahl et al.}(2008){Grundahl, F., Clausen, J.~V., Hardis, S., \& Frandsen, S.}}]{Grundahl08}
     {Grundahl, F., Clausen, J.~V., Hardis, S., \& Frandsen, S.}  2008, \aap, 492, 171

\bibitem[{{Hacking} {et~al.}(1999){hacking}, {Lonsdale}, {Gautier}, {Herter}, {et~al.}}]{Hacking99}
{Hacking}, P., {Lonsdale}, C., {Gautier}, T., {et~al.} 1999, ASPC, Vol. 177, 409

\bibitem[{{Harris \& Canterna}(1981){Harris, W.~E., \& Canterna, R.}}]{Harris_Canterna81} {Harris, W.~E., \& Canterna, R.} 1981, \aj, 86, 1332



\bibitem[{{Hekker} {et~al.}(2011a){Hekker}, {Elsworth}, {De~Ridder}, {et~al.}}]{Hekker11a}
{Hekker}, S., {Elsworth}, Y., {De~Ridder}, J., {et~al.} 2011a, \aap, 252, A131

\bibitem[{{Hekker} {et~al.}(2011b){Hekker}, {Basu}, {Stello}, {Kallinger}, {Grundahl}, {Mathur}, {Garci'a}, {Mosser}, {Huber}, {Bedding}, {Szabo'}, {De~Ridder}, {Chaplin}, {Elsworth}, {Hale}, {Christensen-Dalsgaard}, {Gilliland}, {Still}, {McCauliff}, \& {V.~Quintana}}]{Hekker11b}
{Hekker}, S., {Basu}, S., {Stello}, D., {et~al.} 2011b, \aap, 530, A100

\bibitem[{{Hekker et al.}(2013){Hekker, S., Elsworth, Y., Basu, S., Mazumdar, A., Silva Aguirre, V., \& Chaplin, W.~J.}}]{hekker13}
     {Hekker, S., Elsworth, Y., Basu, S., et al.} 2013, \mnras, 434, 1668


\bibitem[{{Hole et al.}(2009){Hole, K.~T., Geller, A.~M., Mathieu, R.~D., Platais, I., Meibom, S., \& Latham, D.~W.}}]{Hole09}
 {Hole, K.~T., Geller, A.~M., Mathieu, R.~D., et al.} 2009, \aj, 138, 159

\bibitem[{{Jeffries et al.}(2013){Jeffries, M.~W., Jr., Sandquist, E.~L., Mathieu, R.~D., Geller, A.~M., Orosz, J.~A., Milliman, K.~E., Brewer, L.~N., Platais, I., Brogaard, K., Grundahl, F., Frandsen, S., Dotter, A., \& Stello, D.}}]{Jeffries13}
     {Jeffries, M.~W., Jr., Sandquist, E.~L., Mathieu, R.~D., et al.} 2013, \aj, 146, 58



\bibitem[{{Kalirai et al.}(2001){Kalirai, J.~S., Richer, H.~B., Fahlman, G.~G., Cuillandre, J.-C., Ventura, P., D'Antona, F., Bertin, E., Marconi, G., \& Durrell, P.~R.}}]{Kalirai01}
 {Kalirai, J.~S., Richer, H.~B., Fahlman, G.~G., et al.} 2001, \aj, 122, 266

\bibitem[{{Kalirai et al.}(2007){Kalirai, J.~S., Bergeron, P., Hansen, B.~M.~S., Kelson, D.~D., Reitzel, D.~B., Rich, R.~M., \& Richer, H.~B.}}]{Kalirai07}
 {Kalirai, J.~S., Bergeron, P., Hansen, B.~M.~S., et al.} 2007, \apj, 671, 748


\bibitem[{{Kallinger et al.}(2010){Kallinger, T., Mosser, B., Hekker, S., Huber, D., Stello,D., Mathur, S., Basu, S., Bedding, T.~R., Chaplin, W.~J., De Ridder, J., Elsworth, Y.~P.,Frandsen, S., Garc{\'{\i}}a, R.~A., Gruberbauer, M., Matthews, J.~M., Borucki, W.~J., Bruntt,H., Christensen-Dalsgaard, J., Gilliland, R.~L., Kjeldsen, H., \& Koch, D.~G.}}]{Kallinger10}
 {Kallinger, T., Mosser, B., Hekker, S., et al.} 2010, \aap, 522, A1

\bibitem[{{Kaluzny}(1990){Kaluzny, J.}}]{Kaluzny90}
 {Kaluzny, J.} 1990, \mnras, 243, 492

\bibitem[{{Kaluzny \& Rucinski}(1995){Kaluzny, J., \& Rucinski, S.~M.}}]{Kaluzny95}
 {Kaluzny, J., \& Rucinski, S.~M.} 1995, \aaps, 114, 1



\bibitem[{{King et al.}(2005){King, I.~R., Bedin, L.~R., Piotto, G., Cassisi, S., \& Anderson, J.}}]{King05}
     {King, I.~R., Bedin, L.~R., Piotto, G., et al.} 2005, \aj, 130, 626

\bibitem[{{Kinman}(1965){Kinman, T.~D.}}]{Kinman65}
 {Kinman,T.~D.} 1965, \apj, 142, 655



\bibitem[{{Kjeldsen \& Bedding}(1995){Kjeldsen, H., \& Bedding, T.~R.}}]{kjeldsen95}
 {Kjeldsen, H., \& Bedding, T.~R.} 1995, \aap, 293, 87

\bibitem[{{Koch} {et~al.}(2010){Koch, D.~G., Borucki, W.~J., Basri, G., Batalha, N.~M., Brown, T.~M., Caldwell, D., Christensen-Dalsgaard, J., Cochran, W.~D., DeVore, E., Dunham, E.~W., Gautier, T.~N., III, Geary, J.~C., Gilliland, R.~L., Gould, A., Jenkins, J., Kondo, Y., Latham, D.~W., Lissauer, J.~J., Marcy, G., Monet, D., Sasselov, D., Boss, A., Brownlee, D., Caldwell, J., Dupree, A.~K., Howell, S.~B., Kjeldsen, H., Meibom, S., Morrison, D., Owen, T., Reitsema, H., Tarter, J., Bryson, S.~T., Dotson, J.~L., Gazis, P., Haas, M.~R., Kolodziejczak, J., Rowe, J.~F., Van Cleve, J.~E., Allen, C., Chandrasekaran, H., Clarke, B.~D., Li, J., Quintana, E.~V., Tenenbaum, P., Twicken, J.~D., \& Wu, H.}}]{Koch10}
{Koch, D.~G., Borucki, W.~J., Basri, G., et~al.} 2010, \apjl, 713, L79




\bibitem[{{Liebert}(1999){Liebert, J.}}]{Liebert99}
 {Liebert, J.} 1999, ASPC, 192, 143


\bibitem[{{Lindoff}(1972){Lindoff, U.}}]{Lindoff72}
 {Lindoff, U.} 1972, \aaps, 7, 497

\bibitem[{{Matthews et al.}(2004){Matthews, J.~M., Kuschnig, R., Guenther, D.~B., Walker, G.~A.~H., Moffat, A.~F.~J., Rucinski, S.~M., Sasselov, D., \& Weiss, W.~W.}}]{Matthews04}
     {Matthews, J.~M., Kuschnig, R., Guenther, D.~B., et al.} 2004, \nat, 430, 51

\bibitem[{{McCall}(2004){McCall, M.~L.}}]{McCall04}
 {McCall, M.~L.} 2004, \aj, 128, 2144

\bibitem[{{Miglio et al.}(2012){Miglio, A., Brogaard, K., Stello, D., Chaplin, W.~J., D'Antona, F., Montalb{\'a}n, J., Basu, S., Bressan, A., Grundahl, F., Pinsonneault, M., Serenelli, A.~M., Elsworth, Y., Hekker, S., Kallinger, T., Mosser, B., Ventura, P., Bonanno, A., Noels, A., Silva Aguirre, V., Szabo, R., Li, J., McCauliff, S., Middour, C.~K., \& Kjeldsen, H.}}]{Miglio12}
 {Miglio, A., Brogaard, K., Stello, D., et al.} 2012, \mnras, 419, 2077

\bibitem[{{Mochejska et al.}(2003){Mochejska, B.~J., Stanek, K.~Z., \& Kaluzny, J.}}]{Mochejska03}
     {Mochejska, B.~J., Stanek, K.~Z., \& Kaluzny, J.} 2003, \aj, 125, 3175

\bibitem[{{Montgomery et al.}(1994){Montgomery, K.~A., Janes, K.~A., \& Phelps, R.~L.}}]{Montgomery94}
     {Montgomery, K.~A., Janes, K.~A., \& Phelps, R.~L.} 1994, \aj, 108, 585


\bibitem[{{Mosser et al.}(2013){Mosser, B., Michel, E., Belkacem, K., Goupil, M.~J., Baglin, A., Barban, C., Provost, J., Samadi, R., Auvergne, M., \& Catala, C.}}]{Mosser13}
     {Mosser, B., Michel, E., Belkacem, K., et al.} 2013, \aap, 550, A126



\bibitem[{{Perryman \& ESA}(1997){Perryman, M.~A.~C., \& ESA}}]{Perryman_ESA97}
 {Perryman, M.~A.~C., \& ESA} 1997, ESA Special Publication, 1200




\bibitem[{{Ram\'{i}rez} \& {Mel\'{e}ndez}(2005){Ram\'{i}rez}, \& {Mel\'{e}ndez}}]{rm05}
{Ram\'{i}rez}, I., \& {Mel\'{e}ndez}, J. 2005, \apj, 626, 465 (RM05)

\bibitem[{{Rosvick \& Vandenberg}(1998){Rosvick, J.~M., \& Vandenberg, D.~A.}}]{rv98}
 {Rosvick, J.~M., \& Vandenberg, D.~A.} 1998, \aj, 115, 1516



\bibitem[{{Savage \& Mathis}(1979){Savage, B.~D., \& Mathis, J.~S.}}]{SM1979}
    {Savage, B.~D., \& Mathis, J.~S.} 1979, \araa, 17, 73


\bibitem[{{Skrutskie et al.}(2006){Skrutskie, M.~F., Cutri, R.~M., Stiening, R., Weinberg, M.~D., Schneider, S., Carpenter, J.~M., Beichman, C., Capps, R., Chester, T., Elias, J., Huchra, J., Liebert, J., Lonsdale, C., Monet, D.~G., Price, S., Seitzer, P., Jarrett, T., Kirkpatrick, J.~D., Gizis, J.~E., Howard, E., Evans, T., Fowler, J., Fullmer, L., Hurt, R., Light, R., Kopan, E.~L., Marsh, K.~A., McCallon, H.~L., Tam, R., Van Dyk, S., \& Wheelock, S.}}]{Skrutskie06}
     {Skrutskie, M.~F., Cutri, R.~M., Stiening, R., et al.} 2006, \aj, 131, 1163

\bibitem[{{Soszynski et al.}(2008){Soszynski, I., Poleski, R., Udalski, A., Szymanski, M.~K., Kubiak, M., Pietrzynski, G., Wyrzykowski, L., Szewczyk, O., \& Ulaczyk, K.}}]{Soszynski08}
     {Soszynski, I., Poleski, R., Udalski, A., et al.} 2008, \textit{Acta Astronomica}, 58, 163

\bibitem[{{Soszynski et al.}(2010){Soszynski, I., Poleski, R., Udalski, A., Szymanski, M.~K., Kubiak, M., Pietrzy{\~n}ski, G., Wyrzykowski, L., Szewczyk, O., \& Ulaczyk, K.}}]{Soszynski10}
     {Soszynski, I., Poleski, R., Udalski, A., et al.} 2010, \textit{Acta Astronomica}, 60, 17


\bibitem[{{Sandquist et al.}(2013){Sandquist, E.~L., Mathieu, R.~D., Brogaard, K., Meibom, S., Geller, A.~M., Orosz, J.~A., Milliman, K.~E., Jeffries, M.~W., Jr., Brewer, L.~N., Platais, I., Grundahl, F., Bruntt, H., Frandsen, S., \& Stello, D.}}]{Sandquist13}
     {Sandquist, E.~L., Mathieu, R.~D., Brogaard, K., et al.} 2013, \apj, 762, 58




\bibitem[{{Stello et al.}(2008){Stello, D., Bruntt, H., Preston, H., \& Buzasi, D.}}]{Stello08}
 {Stello, D., Bruntt, H., Preston, H., \& Buzasi, D.} 2008, \apjl, 674, L53



\bibitem[{{Stello} {et~al.}(2010){Stello, D., Basu, S., Bruntt, H., Mosser, B., Stevens, I.~R., Brown, T.~M., Christensen-Dalsgaard, J., Gilliland, R.~L., Kjeldsen, H., Arentoft, T., Ballot, J., Barban, C., Bedding, T.~R., Chaplin, W.~J., Elsworth, Y.~P., Garc{\'{\i}}a, R.~A., Goupil, M.-J., Hekker, S., Huber, D., Mathur, S., Meibom, S., Sangaralingam, V., Baldner, C.~S., Belkacem, K., Biazzo, K., Brogaard, K., Su{\'a}rez, J.~C., D'Antona, F., Demarque, P., Esch, L., Gai, N., Grundahl, F., Lebreton, Y., Jiang, B., Jevtic, N., Karoff, C., Miglio, A., Molenda-{\.Z}akowicz, J., Montalb{\'a}n, J., Noels, A., Roca Cort{\'e}s, T., Roxburgh, I.~W., Serenelli, A.~M., Silva Aguirre, V., Sterken, C., Stine, P., Szab{\'o}, R., Weiss, A., Borucki, W.~J., Koch, D., \& Jenkins, J.~M.}}]{Stello10}
{Stello, D., Basu, S., Bruntt, H., et~al.} 2010, \apjl, 713, L182



\bibitem[{{Stello} {et~al.}(2011a){Stello}, {Huber}, {Kallinger}, {Basu}, {et~al.}}]{Stello11a}
{Stello}, D., {Huber}, D., {Kallinger}, T., {et~al.} 2011a, \apjl, 737, L10

\bibitem[{{Stello} {et~al.}(2011b){Stello}, {Meibom}, {Gilliland}, {grundahl}, {Hekker}, {Mosser}, {Kallinger}, {Mathur}, {Garci'a}, {Hubber}, {Basu}, {Bedding}, {Brogaaed}, {Chaplin}, {Elsworth}, {Molenda-Z'akowicz}, {Szabo'}, {Still}, {Jenkins}, {Christensen-Dalsgaard}, {Kjeldsen}, {Serenelli}, \& {Wohler}}]{Stello11b}
{Stello}, D., {Meibom}, S., {Gilliland}, R.~L., {et~al.} 2011b, \apj, 739, 13

\bibitem[{{Stetson et al.}(2003){Stetson, P.~B., Bruntt, H., \& Grundahl, F.}}]{Stetson03}
 {Stetson, P.~B., Bruntt, H., \& Grundahl, F.} 2003, \pasp, 115, 413


\bibitem[{{Talamantes et al.}(2010){Talamantes, A., Sandquist, E.~L., Clem, J.~L., Robb, R.~M., Balam, D.~D., \& Shetrone, M.}}]{Talamantes10}
     {Talamantes, A., Sandquist, E.~L., Clem, J.~L., et al.} 2010, \aj, 140, 1268



\bibitem[{{Torres}(2010){Torres, G.}}]{Torres10}
 {Torres, G.} 2010, \aj, 140, 1158


\bibitem[{{Tripicco et al.}(1995){Tripicco, M.~J., Bell, R.~A., Dorman, B., \& Hufnagel, B.}}]{Tripicco95}
     {Tripicco, M.~J., Bell, R.~A., Dorman, B., \& Hufnagel, B.} 1995, \aj, 109, 1697


\bibitem[{{Walker et al.}(2003){Walker, G., Matthews, J., Kuschnig, R., Johnson, R., Rucinski, S., Pazder, J., Burley, G., Walker, A., Skaret, K., Zee, R., Grocott, S., Carroll, K., Sinclair, P., Sturgeon, D., \& Harron, J.}}]{Walker03}
     {Walker, G., Matthews, J., Kuschnig, R., et al.} 2003, \pasp, 115, 1023

\bibitem[{{Warren \& Cole}(2009){Warren, S.~R., \& Cole, A.~A.}}]{Warren&Cole09}
 {Warren, S.~R., \& Cole, A.~A.} 2009, \mnras, 393, 272

\bibitem[{{Weingartner \& Draine}(2001){Weingartner, J.~C., \& Draine, B.~T.}}]{WD2001} {Weingartner, J.~C., \& Draine, B.~T.} 2001, \apj, 548, 296


\bibitem[{{{White} {et~al.}}(2011){White, T.~R., Bedding, T.~R., Stello, D., Christensen-Dalsgaard, J.,  Huber, D., \& Kjeldsen, H.}}]{White11}
{White, T.~R., Bedding, T.~R., Stell, D., et al.} 2011, \apj, 743, 161

\bibitem[{{{Wu} {et~al.}}(2014){Wu, T., Li, Y., \& Hekker, S.}}]{Wu13}
{Wu, T., Li, Y., \& Hekker, S.} 2014, ApJ, 781, 44

\bibitem[{{Zurek et al.}(1993){Zurek, D.~R., Vandenberg, D.~A., Garnavich,
P.,\& Hesser, J.~E.}}]{Zurek93}
 {Zurek, D.~R., Vandenberg, D.~A., Garnavich, P., \& Hesser, J.~E.} 1993, \jrasc, 87, 217







\end{thebibliography}
\end{document}